\documentclass[useAMS,usenatbib,referee]{mn2e}
\usepackage{graphicx}
\usepackage{amssymb}
\usepackage[T1]{fontenc}
\usepackage{ae,aecompl}
\usepackage{subfig}
\usepackage{threeparttable}
\usepackage{color}

\usepackage{float}
\usepackage{epsfig}
\title[HiRes inner-shell of P Cyg nebula in {\rm [}{\rm Fe\,II}{\rm ]}]{A high resolution image of the inner-shell of the P\,Cygni nebula in the infra-red  {[}Fe\,II{]} line}

\author[C. Arcidiacono et al. ]
{C. Arcidiacono$^{1}$\thanks{E-mail:
carmelo.arcidiacono@oabo.inaf.it (CA)},  R. Ragazzoni$^{2}$, C. Morossi$^{3}$, M. Franchini$^{3}$, P. Di Marcantonio$^{3}$,  \newauthor
C. Kulesa$^{4}$, D. McCarthy$^{4}$, R. Briguglio$^{5}$, M. Xompero$^{5}$, L. Busoni$^{5}$, 
\newauthor
F. Quir$\acute{o}$s-Pacheco$^{5}$, 	E. Pinna$^{5}$, K. Boutsia$^{6}$, and D. Paris$^{6}$\\
$^{1}$INAF -- Osservatorio Astronomico di Bologna, Via Ranzani, 1, I--40127 Bologna, Italy\\
$^{2}$INAF -- Osservatorio Astronomico di Padova, Vicolo dell'osservatorio, 5, I--35122 Padova, Italy\\
$^{3}$INAF - Osservatorio Astronomico di Trieste, Via G. B. Tiepolo 11, Trieste, I--34143, Italy\\
$^{4}$Steward Observatory Annex, 1540 East Second Street Tucson, AZ 85721-0064, USA\\
$^{5}$INAF -- Osservatorio Astrofisico di Arcetri, Largo Enrico Fermi, 5, I--50125 Firenze, Italy\\
$^{6}$INAF -- Osservatorio Astronomico di Roma, Via Frascati, 33, I--00040 Monte Porzio Catone, Italy}

\begin{document}

\date{Accepted . Received; in original form }

\pagerange{\pageref{firstpage}--\pageref{lastpage}} \pubyear{2014}

\maketitle

\label{firstpage}

\begin{abstract}
We have obtained  with the LBT Telescope AO system Near-Infrared camera PISCES images of the  inner--shell of the nebula around the luminous blue 
variable star P Cygni in the [Fe\,II] emission line  at 1.6435\,$\mu$m. We have combined the images in order to
cover a field of view of about 20$\arcsec$ around P Cygni thus providing  the high resolution ($0\arcsec.08$) 2--D spatial distribution 
of the  inner--shell of the P Cygni nebula in [Fe\,II].
We have identified several nebular emission regions  which are characterized by an S/N$>$3. 
A comparison of our results with those available in the literature shows  full consistency 
with the finding by \citet{SM06} which are based on radial velocity measurements
and their relatively good agreement with the extension of emission nebula in [N\,II] $\lambda$6584 found by \citet{BA94}.
We have clearly detected extended emission also inside the radial distance $R$=7$\arcsec$.8 and outside  $R$=9$\arcsec$.7
which are the nebular boundaries proposed by  \citet{SM06}. New complementary spectroscopic observations 
to measure radial velocities and to derive  the 3--D distribution of P Cygni nebula are planned.

\end{abstract}


\section{Introduction}
The star P Cygni was discovered on August 18, 1600 AD when it brightened
to about the third magnitude. It is a ``prototypical'' Galactic Luminous Blue Variable (LBV) star
and its basic properties are described by \citet{NA97} and \citet{NA01}.

Several attempts to get informations on size, chemical composition and density 
distribution of the material around the star have been already done (see
for example \citealt{LA87}). \citet{BA94}  got
[N\,II] $\lambda$6584 narrow band images showing  unresolved clumps of emission 
distributed within a  nearly circular nebula with
angular diameter of about 22$\arcsec$ and hint at the presence of  
nebular rings with radii of  6$\arcsec$ and 11$\arcsec$. \citet{BA94}
found clumps, typically 2$\arcsec$--3$\arcsec$ in diameter, not symmetrically 
distributed around the star (see \citealt{NO95}).
On the other hand, \citet{WB82} observations are consistent
with a homogeneous, spherically symmetric, constant velocity, isothermal flow. 
An outer shell with a radius of $\sim$1\arcmin.6 is also present indicating mass ejection
before 1600\,AD (see for example \citealt{ME96}, \citealt{ME99}, \citealt{ME04})

Near-InfraRed (NIR) spectroscopy of P Cygni (\citealt{SM01} and \citealt{SM06}) revealed bright emission
in the [Fe\,II] line at $\lambda$=1.6435\,$\mu$m.
\citet{SM06} derived from the analysis of high--resolution spectra the following geometric 
parameters of P\,Cygni inner--shell: a mean expansion velocity of 136\,km\,s$^{-1}$, a shell radius
of about 10$^{17.3}$\,cm, a dynamic age of about 530\,yr, and a geometric filling factor in [Fe\,II]
roughly equal to 0.2. They concluded that the [Fe\,II] emitting shell at about 10$\arcsec$ from P Cygni is
the product of the 1600\,AD outburst.  In their results the cause of greater uncertainty was the uncertainty in 
the adopted P\,Cygni distance (1.7$\pm$0.1\,kpc) and the need of assuming a filling factor. 
In fact, the choice of the latter was somehow arbitrary since they had no NIR images  and only relatively 
low spatial resolutions orthogonally and along their long slit spectra, i.e. $\sim0\arcsec.5$ and $\sim2\arcsec$, respectively.

A significant step forward in the knowledge of P\,Cygni nebula in particular and of LBV
stars in general would require high resolution 3--D information on the nebular structure which,
in our opinion,  has  become possible with the availability, nowadays, of large ground--based telescopes, 
technologically advanced instruments, and   Adaptive Optics (AO) systems.
In this paper we present a collage of 72 Near-Infrared camera PISCES \citep{MC01} images obtained by the Large Binocular Telescope (LBT, 
\citealt{HI10})\footnote{The LBT is an international 
collaboration among institutions in the United States, 
Italy and Germany. LBT Corporation partners are: The University of Arizona on behalf of the Arizona university system; 
Istituto Nazionale di Astrofisica, Italy; 
LBT Beteiligungsgesellschaft, Germany, representing the Max-Planck Society,
the Astrophysical Institute Potsdam, and Heidelberg University; The Ohio State
University, and The Research Corporation, on behalf of The University of Notre Dame, 
University of Minnesota, and University of Virginia.} AO system in a narrow band 
filter centered on the 1.6435\,$\mu$m NIR forbidden line of Fe\,II. The final image covers a field of view of about 20$\arcsec$ around
P\,Cygni and provides high resolution 2--D information on P\,Cygni nebula which will be complemented with  future
radial velocity observations in order to derive the nebular distribution along the line of sight.
In Section \ref{obs} we present the main characteristics
of the instrumental setup and of the acquired AO images; in  Section \ref{reduction}
we describe the image processing which lead to a composite image of the nebular emission around P\,Cygni from 
the set of 72 individual images;
in Section~\ref{result} the 2--D spatial distribution of P\,Cygni nebula is presented and our results are compared
with those already available in the literature.

\section{Observations}
\label{obs}

\subsection{The Adaptive Optics}

Both telescopes of the LBT  are equipped with a high order adaptive optics system called First Light Adaptive Optics, FLAO \citep{ES12}. 
The FLAO senses the atmospheric turbulence by means of a pyramid based wavefront sensor \citep{RA96} which samples the pupil on a grid of  $30\times30$ sub-apertures.
A corrected wavefront is obtained through an Adaptive Secondary Mirror (ASM) (\citealt{SA93} and \citealt{RI10}) 
controlled by 672 actuators at a frequency of 1\,kHz. 
The spatial and temporal configurations of the FLAO match the performance of the ASM. 
On typical seeing values of 0\arcsec.8 the FLAO provides in the H band high order corrected images with Strehl Ratios of about 80\% on bright reference sources.

P\,Cygni is a bright reference for the FLAO system, whose R band magnitude operative range is between 3.5 and the 17\,mag. 
Actually, the FLAO system, whose filter of the wavefront sensor is centered  at 0.7$\mu$m,  detected this star as  a 5.5\,mag object.

\subsection{PISCES Camera}
\label{pisces}
In order to provide early science with the LBT Adaptive Optics system before the arrival at the telescope of the diffraction limited camera, 
we have the opportunity to utilize the PISCES infrared camera provided by the University of Arizona.  PISCES provides  diffraction limited images 
in the 1--2.5\,$\mu$m range with a pixelscale of 0.019\,arcsec/px. Further details on PISCES camera can be found
in \cite{MC01}, \cite{ES13} and \cite{GU12}.

\subsection{Observing strategy}
The observations of P\,Cygni were obtained on June 22$^{\rm nd}$ 2012 in the [FeII] emission line filter centered on the emission line at 1.6435\,$\mu$m.
 In order to avoid saturation and to cover a field of about 46\arcsec{} in diameter centered on P\,Cygni, 
we intentionally moved the star out of the field of view. We used 12 different observational sets:
the first four were obtained by putting P\,Cygni about 0\arcsec.5 North, South, East, or West out of the field of view; the other eight set were obtained
from the first four after applying a 30 or 60 degrees rotation of the LBT Gregorian Rotator. We took six 20\,s images in each  observational set
leading to a total of 72 images. During the observations the Differential Image Motion Monitor (DIMM) on board the LBT measured a variable 
seeing between 0\arcsec.9 to 1\arcsec.5  (V band). 
The telescope elevation was ranging from 70 to 84 degrees and the DIMM was pointing as close as possible to the target.

\section{Data Reduction}
\label{reduction}
We follow a standard data reduction method for near infrared data. Electronic cross--talk between the quadrants in the PISCES detector
was corrected using  ``Corquad'', an IRAF\footnote{IRAF  2011, ``Image Reduction and Analysis Facility'', Version 2.15.1a, NOAO, Tucson}
 procedure developed by Roelof de Jong\footnote{See website http://66.194.178.32/~rfinn/pisces.html)}. 
The Full Width at Half Maximum (FWHM) of a few stars present in each reduced frame is about 0\arcsec.07 -.10\arcsec, slightly variable along the field of view
 because of the anisoplanatism effect. The average FWHM and Strehl Ratio (SR) are 0\arcsec.08 and  12\%, respectively.
 
\begin{table*}
\begin{threeparttable}
%\small
 \centering
% \begin{minipage}{140mm}
  \caption{P\,Cygni nebula emission regions}
  \label{tab1}
  \begin{tabular}{@{}crrcllc@{}}
  \hline
   \multicolumn{1}{c}{Region} & Size & Tot. Int.  &  \multicolumn{1}{c}{Knot \#}  &  RA &  \multicolumn{1}{c}{DEC} &  \multicolumn{1}{c}{Max Int.}  \\
        &  arcsec$^2$  & counts &  &  \multicolumn{1}{c}{hh:mm:ss}  & \multicolumn{1}{c}{$\deg:\min:\sec$} & counts pix$^{-1}$ \\
  \hline
 A & 7.4 & 62331 & & &&\\
B & 1.4 & 19777 & 1 & 20:17:47.7  & +38:02:13.9 & 204\tnote{a}\\
C & 1.0 & 8691 & &  &&\\
D & 4.8 & 39952 &  &    &  &  \\
E & 0.4 & 3809 &  &    &  &  \\
F & 4.2 & 48335 &  &    &  &  \\
G & 16.3 & 244982 & 1 & 20:17:47.9   & +38:01:59.1  & 17 \\
 & &  & 2 &  20:17:47.7 & +38:01:56.0  & 15 \\
H & 2.3 & 20145 &  &    &  &  \\
I & 1.6 & 11607 &  &    &  &  \\
J & 4.8 & 49098 & 1 & 20:17:47.6   &+38:01:51.2  & 13 \\
K & 4.6 & 41238 & 1 & 20:17:46.6   &+38:01:41.0  & 126\tnote{a}\\
L & 3.3 & 29458 &  &    &  &  \\
M & 12.0 & 115791 &  &    &  &  \\
N & 1.4 & 10499 &  &    &  &  \\
O & 2.4 & 17964 &  &    &  &  \\
P & 4.6 & 34151 &  &    &  &  \\
Q & 0.8 & 7509 &  &    &  &  \\
R & 197.9 & 2.99E6 & 1 &  20:17:46.7 & +38:02:00.3 & 277\tnote{a} \\
 & &  & 2 &  20:17:46.9 & +38:02:01.9 & 29 \\
 & &  & 3 & 20:17:47.2   & +38:02:04.7  & 28 \\
 & &  & 4 &  20:17:46.9 & +38:02:02.2 & 23 \\
 & &  & 5 &  20:17:47.1 & +38:02:06.6 & 21 \\
 & &  & 6 &  20:17:47.1 & +38:02:04.2 & 21 \\
 & &  & 7 &  20:17:47.1 & +38:02:06.5 & 20 \\
 & &  & 8 &  20:17:47.3 & +38:02:03.7 & 20 \\
 & &  & 9 &  20:17:47.0 & +38:02:05.8 & 19 \\
 & &  & 10 &  20:17:47.0 & +38:01:52.7 & 18 \\
 & &  & 11 &  20:17:46.7 & +38:02:05.1 & 14 \\
 & &  & 12 &  20:17:47.6 & +38:02:03.3 & 14 \\
 & &  & 13 &  20:17:47.5 & +38:02:04.4 & 14 \\
 & &  & 14 &  20:17:47.4 & +38:02:06.8 & 13 \\
 & &  & 15 &  20:17:46.6 & +38:01:50.3 & 12 \\
 & &  & 16 &  20:17:46.9 & +38:02:04.7 & 12 \\
 & &  & 17 &  20:17:46.7 & +38:02:04.2 & 11 \\
 & &  & 18 &  20:17:46.6 & +38:01:51.8 & 10 \\ 
\hline
\end{tabular}
%\footnotetext[1]{very likely a background source}
\begin{tablenotes}
 \item[a] Very likely a background sources
\end{tablenotes}

%\end{minipage}
\end{threeparttable}
\end{table*}

\subsection{Astrometry}
\label{sec:sstrometry}
\begin{figure*}
\hspace*{-2cm}
\vspace*{-2cm}
\includegraphics[width=1.2\textwidth]{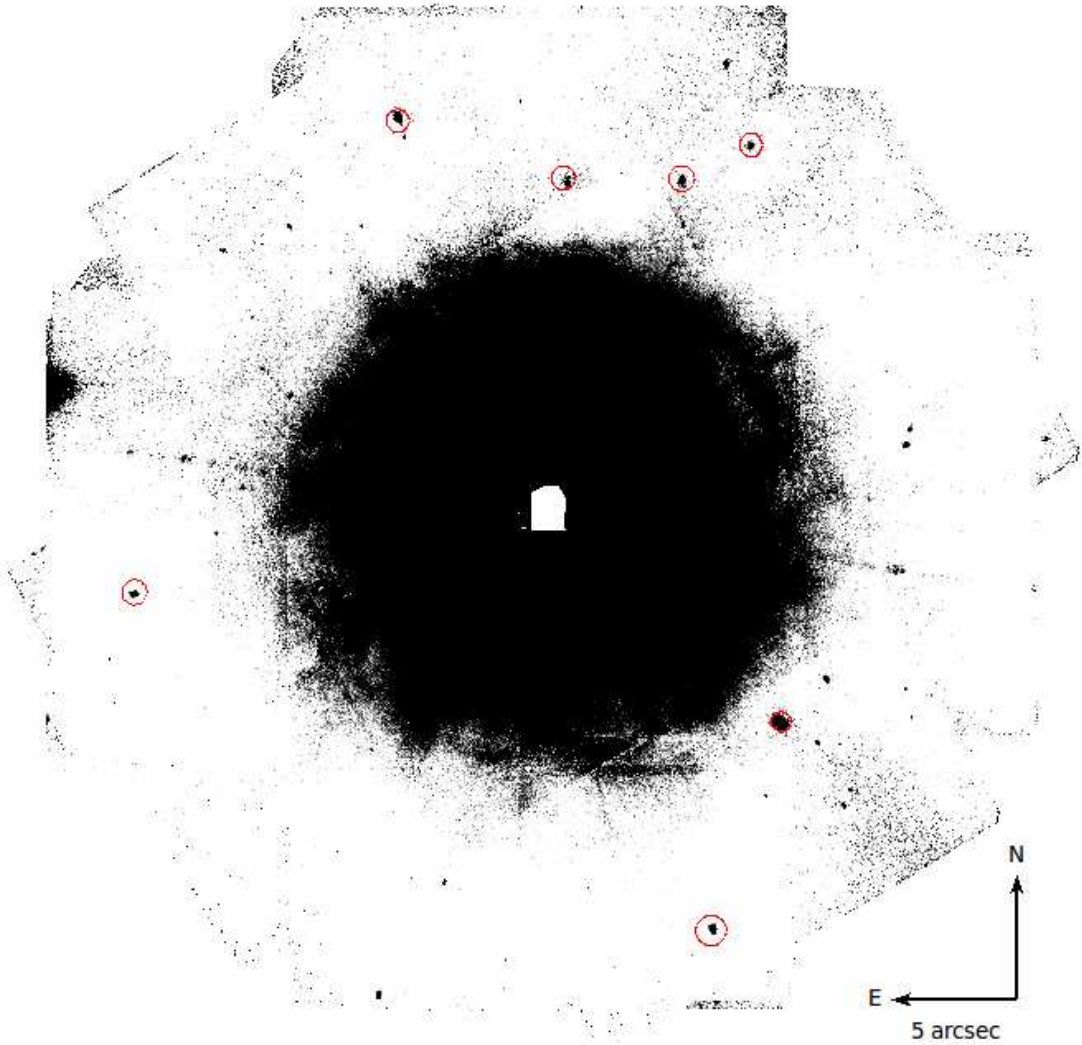}
\caption{ {\it Comb} image   in the [Fe\,II] nebular emission line of the inner--shell of the P Cygni nebula. The superimposed red circles are centered on the 
 centroid positions of the comparison stars in the WFPC2 image.} 
\label{comb_im}
\end{figure*}

To improve the transformation from image coordinates in pixel to astronomical RA and DEC we searched for
astrometrically calibrated images in the available databases and we found several Wide Field and Planetary Camera 2 (WFPC2) images taken with
F658N filter. Then, we identified on the image built by co-adding the retrieved images several faint stars
surrounding P\,Cygni. The same stars were searched on our images and their positions were used in {\it ccmap}
and in {\it ccsetwcs} IRAF tasks
 to add a World Coordinate System (WCS) to each of our individual frames. Eventually, the 72 frames were combined with the {\it imcombine} IRAF
task to create a merged/collaged {\it Comb} image and a $\sigma$--image covering about an almost circular field of view
of about 23\arcsec{} in radius with a central vignetting of about 1\,\arcsec. 
Both images have astrometric accuracy  on the order of $\pm0\arcsec.15$
as derived by the residuals of the position fit in {\it ccmap}. The   {\it Comb} image
is shown, in arbitrary intensity scale,  in Fig.\,\ref{comb_im}   with 
superimposed circles marking the positions of the comparison stars derived from the WFPC2 image. It is clearly evident that the
scatter light of the P\,Cygni Point Spread Function (PSF) dominates and must be removed before investigating the nebular emission.
As far as the angular resolution of the  {\it Comb} image is concerned the analysis of the comparison stars
indicate an average FWHM of about 0$\arcsec$.08 confirming that the  LBT AO system was able to achieve diffraction limit quality 
in the [Fe\,II] $\lambda$=1.6435\,$\mu$m filter.

\begin{figure*}
\hspace*{-1cm}
\includegraphics[width=0.8\textwidth,angle=90]{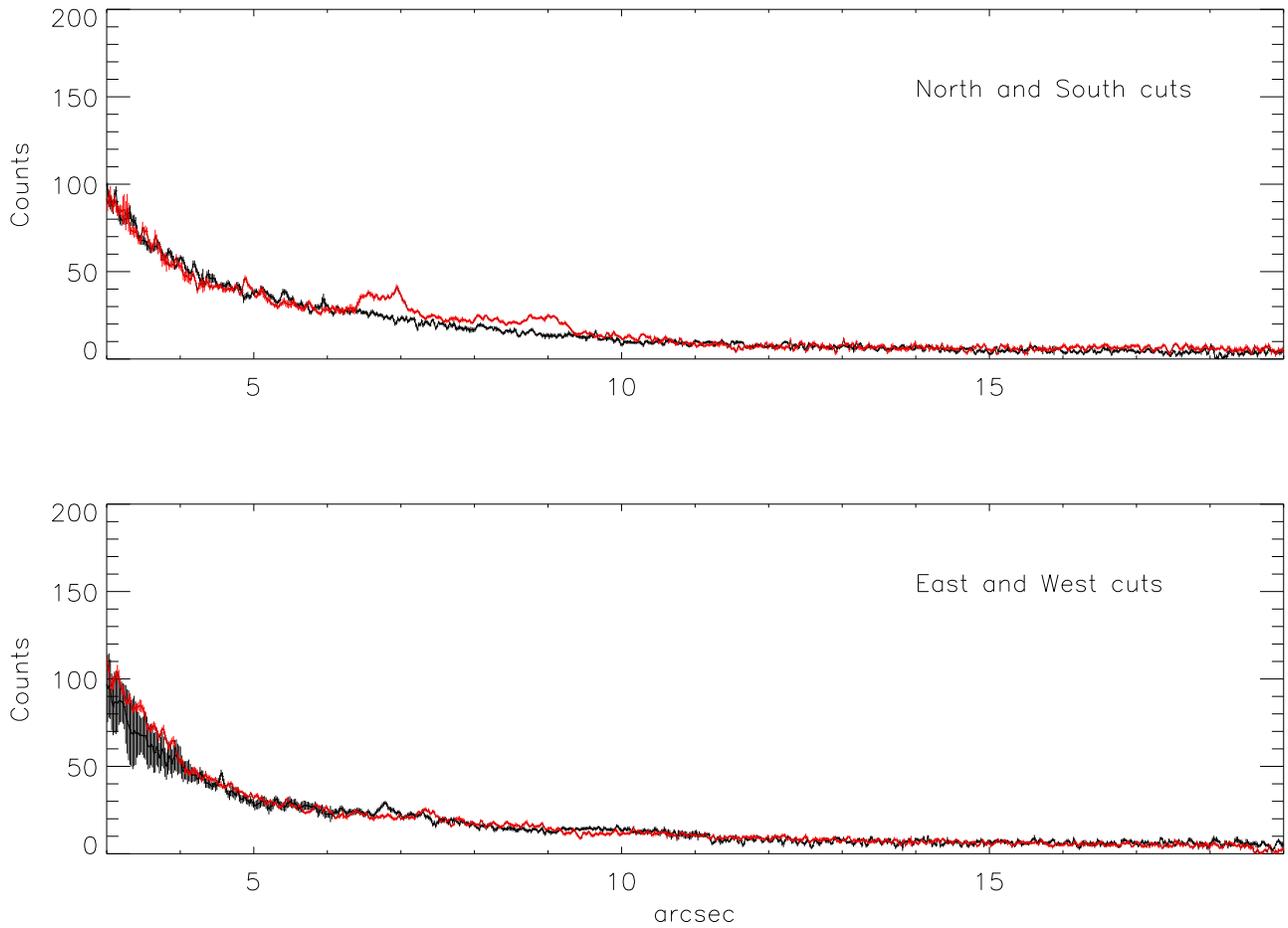}
\caption{ {\it Comb} image cuts starting at nominal P\,Cygni position. Upper panel: towards South (black, $\theta=\pi$) and towards North (red, $\theta=0$); lower panel: 
towards East (black, $\theta=3/2\pi$) and towards West (red, $\theta=\pi/2$). The excesses of counts at distances of about 
$R=7\arcsec$ and $R=$9$\arcsec$ from P\,Cygni position in the red cut of the upper panel with respect to the black one are due to nebula emission regions 
(see Fig.\,\ref{neb_reg} and Sec.\,\ref{result}) }
\label{prof}
\end{figure*}

Assuming a circular symmetric PSF, we computed radial profiles of P\,Cygni scatter light distribution at
different azimuth angles $\theta_{\rm i}$. The hypothesis of circular symmetry was then checked by over-plotting pairs of radial profiles
at $\theta_{\rm i}$ and $\theta_{\rm i}+\pi$ (see some examples in Fig.\,\ref{prof}).
Finally, the distributions of intensity of the radial profiles at different distances, $R$, from the nominal position of 
P\,Cygni were built and analyzed leading to, for each $R$, at the determination of the intensity of the mean angular
radial profile of the scattered light. To do this, we assumed that the minimum (after removing a few clear outliers)
of the distribution of intensity at each radial distance is not affected by a significant nebular emission. 
We were able to derive a reliable radial profile only in an annulus around P\,Cygni 
position characterized by an inner and an outer radius of $R=$3$\arcsec$ and $R=$19$\arcsec$ respectively.
In fact, at $R<$3$\arcsec$ there is practically no pair ($\theta_{\rm i}$, $R$) where a contribution of nebular emission
can be excluded {\it a priori},
while at $R>$19$\arcsec$ the radial profiles are very noisy due to the reduced number of overlapping individual frames
 used to obtain the {\it Comb} image at these distances from P Cygni and to the intrinsic low signal in each individual frame. 

\begin{figure*}
\hspace*{-1cm}
\vspace*{-1.5cm}
\includegraphics[width=1.25\textwidth]{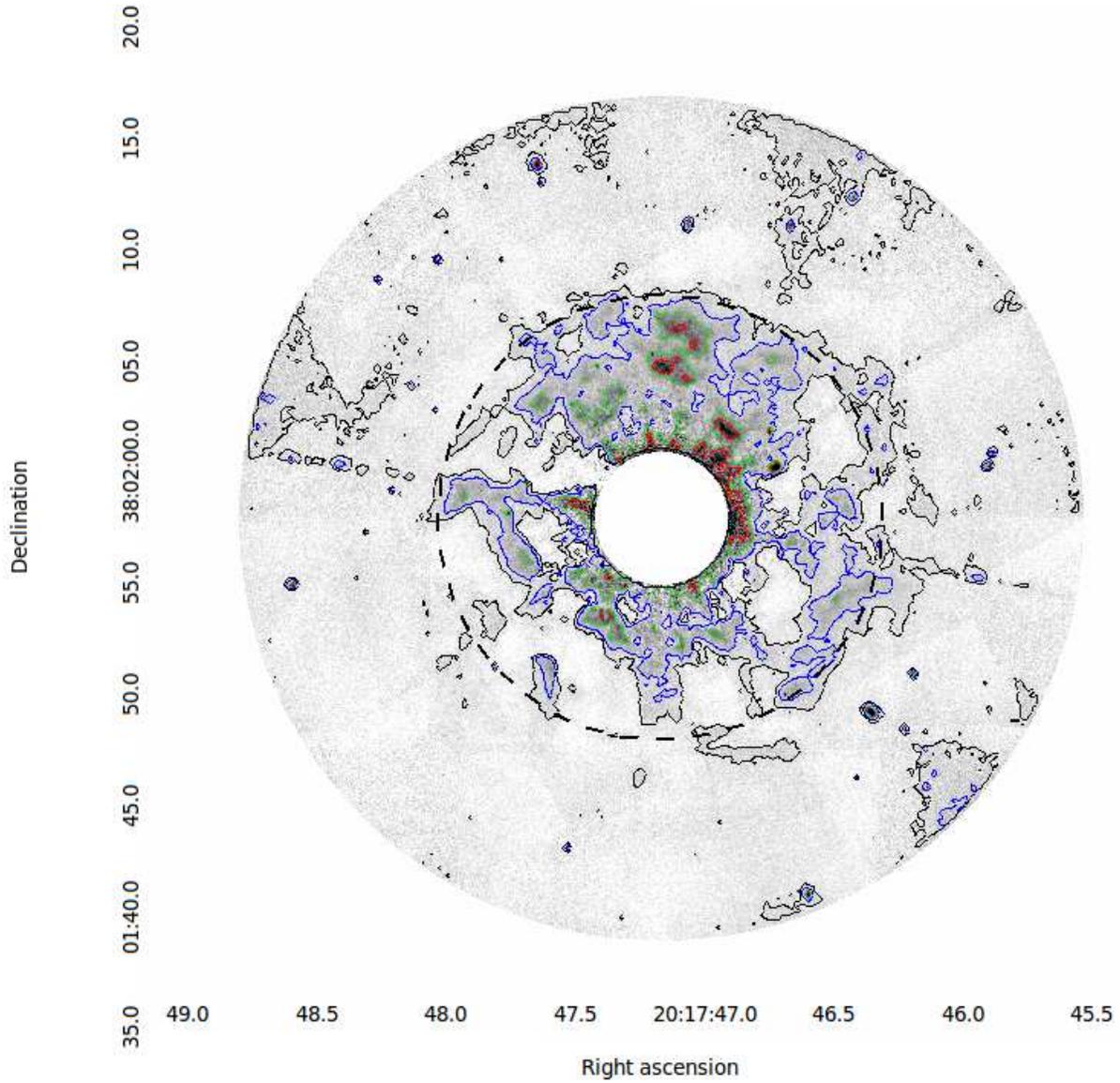}
\caption{ {\it Neb} image of the inner--shell  of the P\,Cygni nebula, obtained by subtracting from {\it Comb} image a model of P\,Cygni scattered light built using an average radial profile (see text), with 
superimposed intensity contour levels at 3 (black), 5 (blue), 10 (green), and 15 (red) counts. The radius of the dashed black circle is 10$\arcsec$. }
\label{neb_im}
\end{figure*}

The image of the nebular emission, {\it Neb}, was eventually obtained by subtracting the above--described 
 scattered light model from the {\it Comb} image and it is shown in Fig.\,\ref{neb_im}.

\section{ Results}
\label{result}
\begin{figure*}
\hspace*{-1cm}
\vspace*{-1.5cm}
\includegraphics[width=1.25\textwidth]{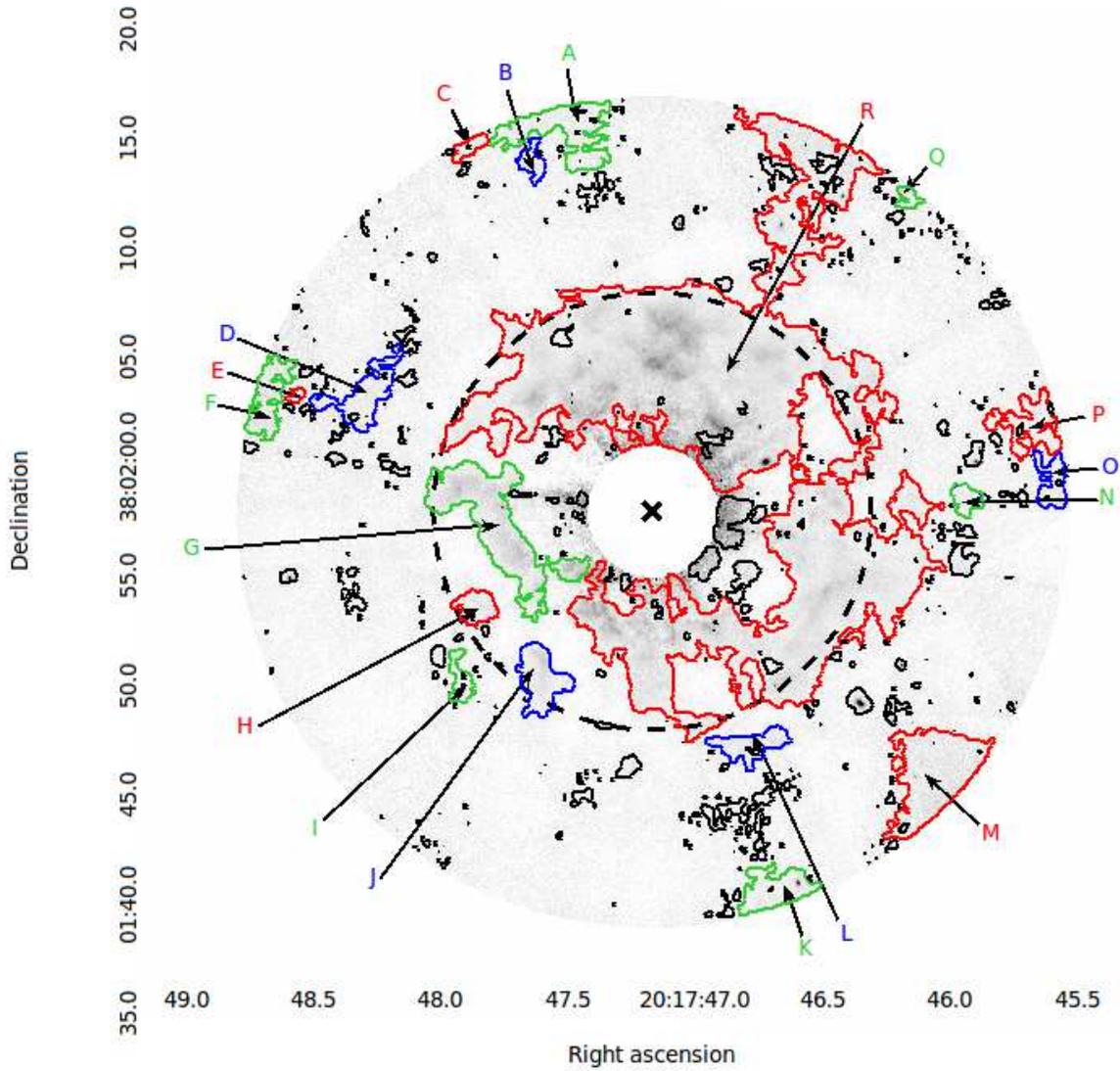}
\caption{Most prominent   [Fe\,II] emission regions in  the inner--shell of the P\,Cygni nebula defined as closed areas within S/N=3 boundary. Main parameters of each area identified
by an alphabetic character can be found in Table\,\ref{tab1}. The position of P\,Cygni is indicated by ``X'' symbol, the radius of the dashed black circle is 10$\arcsec$. }
\label{neb_reg}
\end{figure*}

By comparing the intensity levels of the {\it Neb} image with those of the $\sigma$--image, we were able to identify 
in the  former several nebular emission regions  in the [Fe\,II] line at $\lambda$=1.6435\,$\mu$m (labeled in Fig.\,\ref{neb_reg})
which can be defined as those close regions characterized by an Signal--to--Noise Ratio (S/N)  greater than 3. 
In Table\,\ref{tab1} we report, for each region, its size and  total intensity within its boundary defined by the S/N=3 threshold,
and the positions and the intensity of its emission maxima (knots).
As can be seen in  Fig.\,\ref{neb_reg} most of the emission regions are consistent with the finding by \citet{SM06}
which stated, on the bases of radial velocity measurements, that  inner--shell of the P\,Cygni nebula can be approximated with a spherical shell
characterized by an inner an outer radius at a distance of $R=7\arcsec.8$ and $R=9\arcsec.7$, respectively. 
Our results are also in relatively good agreement  with the image in [N\,II] $\lambda$6584 of the inner--shell  found by
\citet{BA94} even if we do not find a larger diameter in the North--South direction than in the East--West direction. 
On the other hand, our results show  
clearly the presence of extended emission also inside $R=7\arcsec.8$ and outside  $R=9\arcsec.7$  
with some asymmetry due to stronger emission at North--Northwest than at South--East of P\,Cygni. 
There are also hints of emission regions at larger radial distances, i.e. $R > 16\arcsec$ and of some possible connection 
with the internal regions in the North--West (see region R in Fig.\,\ref{neb_reg}). While we are  very  confident on the results
we obtained for $R\le15\arcsec$ we must point out that for larger $R$ we have the co--addition of only few individual frames 
and larger uncertainties on the subtracted average radial profile of P\,Cygni scattered light. Thus, the presence of a second
region of nebular emission at the border of our field of view would require more dedicated observations to be soundly confirmed.

\begin{figure*}
\hspace*{-1cm}
\vspace*{-1.5cm}
\includegraphics[width=1.25\textwidth]{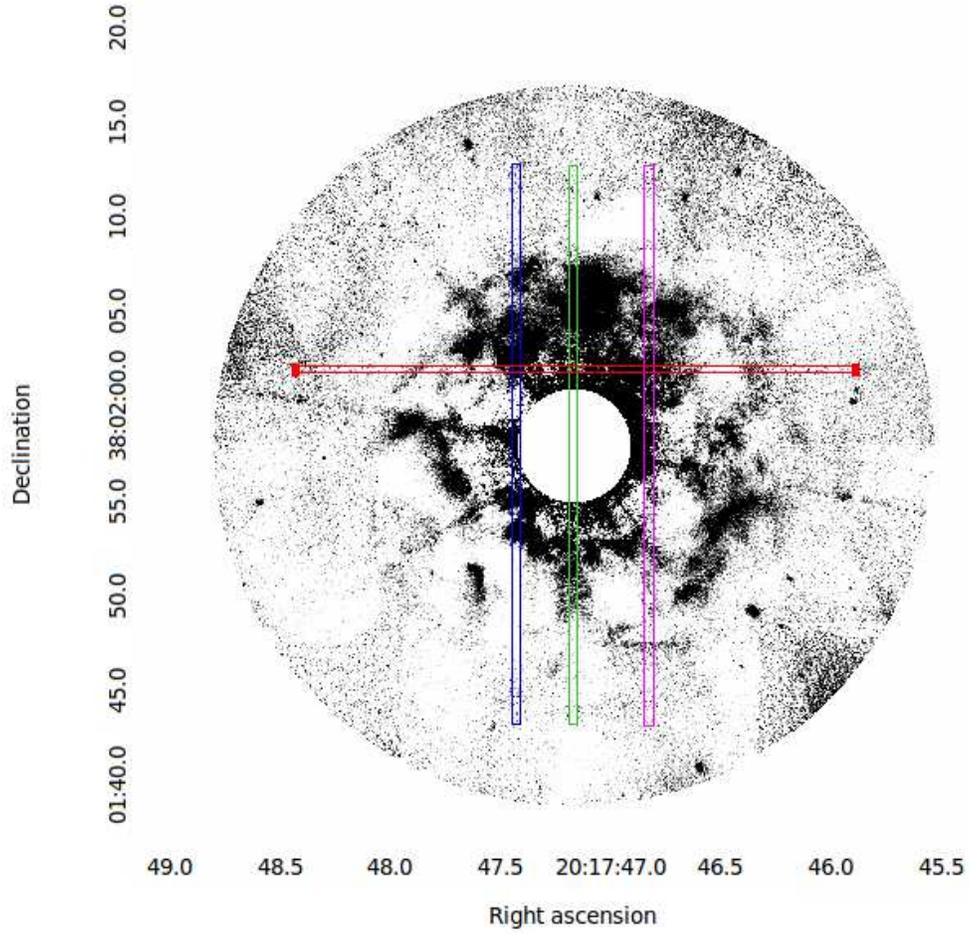}
\caption{ Long slits used in the paper by \citet{SM06} superimposed over {\it Neb} [Fe\,II] image of the  inner--shell of P\,Cygni nebula (see text). }
\label{slits}
\end{figure*}

A direct comparison of our results with those already reported in the literature can be done by superimposing over the {\it Neb} image
the {\it Spex} (red box in Fig.\,\ref{slits}) and the three {\it CSHELL} long slits (blue, green, and magenta boxes in Fig.\,\ref{slits})
used in the paper by \citet{SM06}. 

\begin{figure*}
\hspace*{-2cm}
\includegraphics[width=0.8\textwidth,angle=90]{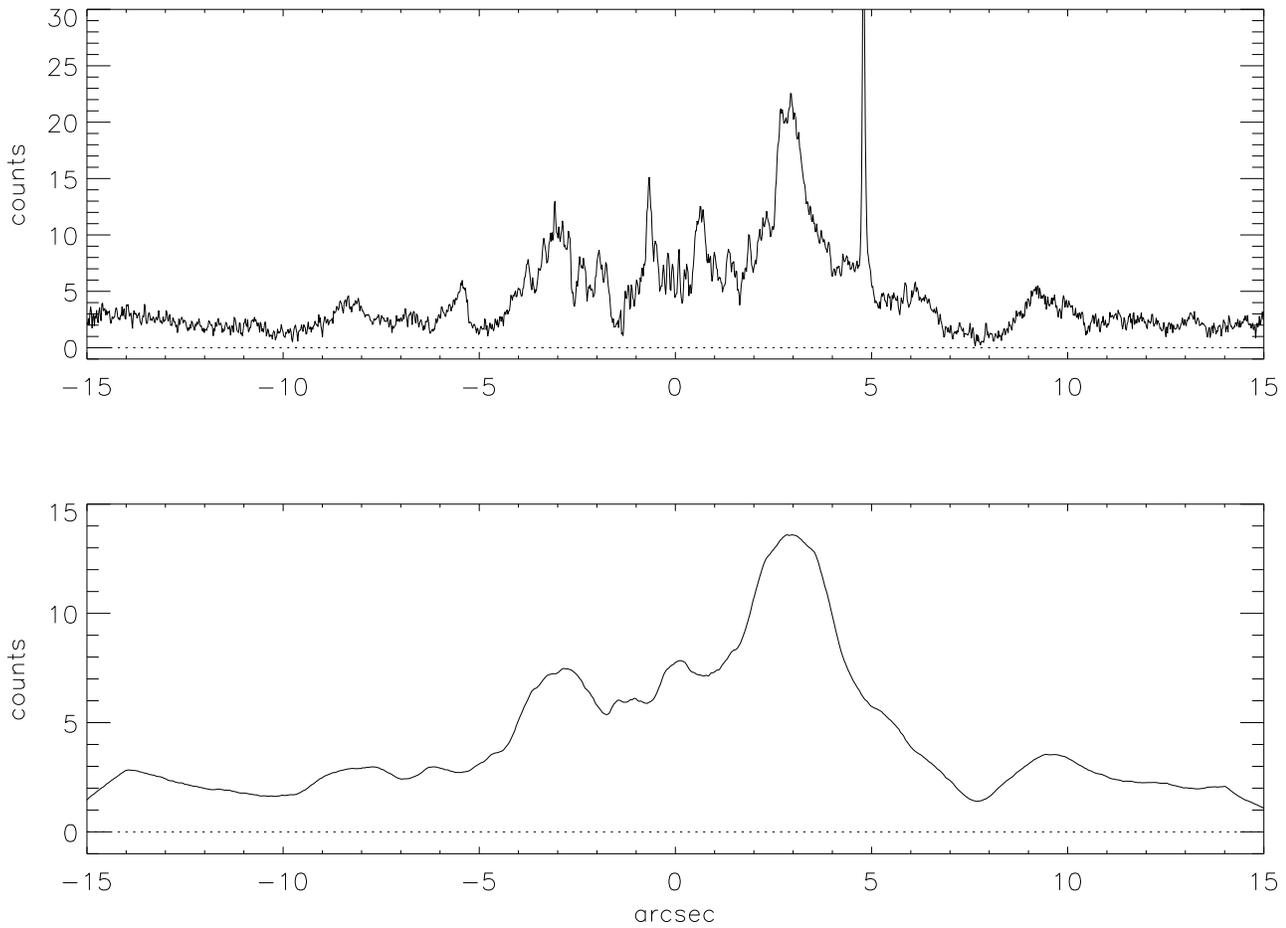}
\caption{ [Fe\,II] emission intensity in the East-West area (red box in Fig.\ref{slits}) 0$\arcsec$.3 wide centered at 4$\arcsec$ north of the inner--shell
 of the P\,Cygni nebula. Upper panel:  the original data in the [Fe\,II] nebular emission line (the peak at 4$\arcsec$.9 West is due to a
background source); lower panel: original data degraded at 2$\arcsec$ spatial resolution to mimic upper panel of Fig.\,4 by \citet{SM06} after removing
the background source. }
\label{spex}
\end{figure*}

By integrating in RA the {\it Neb} image counts contained in the {\it Spex} slit region we obtained Fig.\,\ref{spex} 
which can be compared with top panel of Fig.\,4 by 
\citet{SM06}. As can be seen, there is a general agreement between our results and those previously reported by
\citet{SM06}.  It is also very clear, by comparing the upper and the lower panel,
the gain in spatial resolution we achieved by taking advantage of the LBT AO system.
It is worthwhile noticing that our integrated counts contains not only the nebular emission at the [Fe\,II] line at $\lambda$=1.6435\,$\mu$m 
but, actually, the whole nebular emission in the filter wavelength band, thus  including also some contribution from the weak nebular continuum. 
This continuous emission, which can be seen in third panel of Fig.\,3 by  \citet{SM06}, 
can explain   why our integrated counts in the lower panel of Fig.\ref{spex} do not go to zero at a distance of
about 15$\arcsec$ East or West from P\,Cygni RA as in Fig.\,4 by \citet{SM06}.

\begin{figure*}
\hspace*{-2cm}
\includegraphics[width=0.8\textwidth,angle=90]{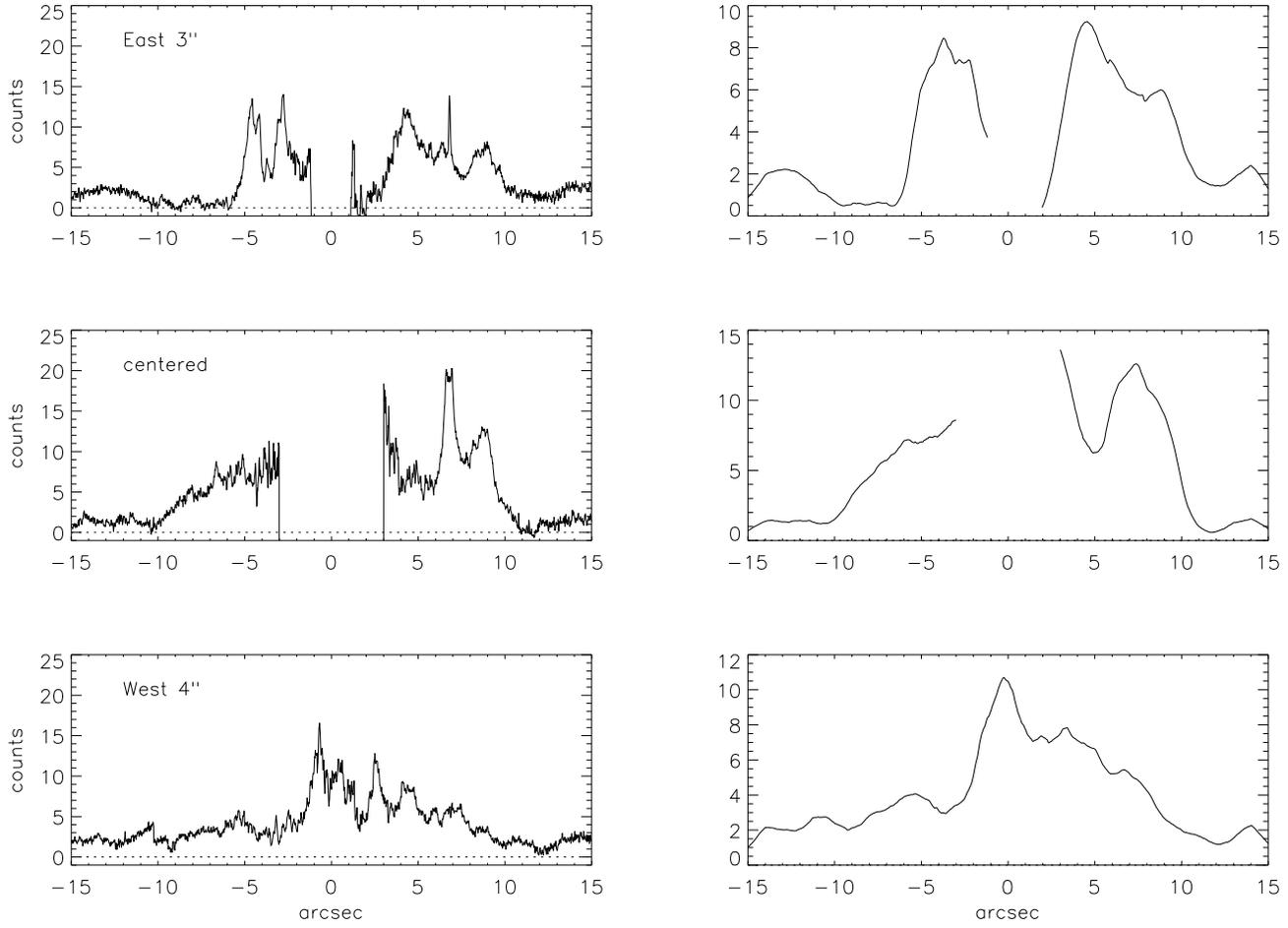}
\caption{ [Fe\,II] emission intensity in the South-North areas (blue, green, and magenta boxes in Fig.\ref{slits}) 0$\arcsec$.5 wide centered at 3$\arcsec$ East
of P\,Cygni (upper panels), on star (middle panels), and at 4$\arcsec$ West of P\,Cygni (lower panels).
Left panels: original data; right panels: original data degraded at 2$\arcsec$ spatial resolution to mimic  Fig.\,5 by \citet{SM06}. }
\label{cshell}
\end{figure*}

As far as the  {\it CSHELL} slits are concerned, the integration in DEC of the {\it Neb} image counts should be compared with
Fig.\,5 by \citet{SM06} taking into account that we do not have radial velocity information. In any case,
there is a qualitative agreement since the maxima in the right panels of Fig.\,\ref{cshell}  fall at the same DEC of the knots in
Fig.\,5 by \citet{SM06}. It is clear that the intensity of the different maxima in Fig.\,\ref{cshell} are the results of co--adding
the emission of nebular regions along the line of sight characterized by different radial velocities.
The same effect is obviously present in Fig.s\,\ref{neb_im}--\ref{spex} showing that to derive the actual 3--D mapping of
the inner--shell of P\,Cygni nebula
we have to wait until complete coverage in both imaging and radial velocity will be available.

\section{ Future observations}
\label{Future}
With the aim of deriving a full 3--D picture of the inner--shell of PCygni nebula, we aim
to obtain NIR spectra in the wavelength region  of the strongest [Fe\,II] nebular lines  
to discriminate in our image the  approaching and the receding parts of the ejecta
by measuring their radial velocities. 
These spectroscopic measurements, together with our 2--D image, will allow us to use a more detailed geometry
and a more constrained filling factor value than those used by \citet{SM06}, 
to derive more reliable and accurate estimate of the ejecta mass.
Furthermore, the radial velocity measurements in the regions at radial distances larger than 10$\arcsec$
might allow us to get hints on the dates of ejection episode(s) that occurred before the 1600 AD outburst. 
In such a way we could estimate the typical time period between such eruptions which is needed in order to understand how
much of the mass loss, which causes a massive stars to become a Wolf--Rayet star, occurs in giant brief eruptions.

In conclusion, it is also worthwhile noticing that it could be feasible to measure the expansion proper motions of the inner--shell of P\,Cygni nebula 
starting from the high spatial resolution (0$\arcsec$.08) image  presented in this paper by obtaining,  in the near future,
new NIR AO images. In fact, assuming a constant expansion velocity of $\sim$140\,km/s
and a distance of 1.7\,kpc, it would be possible to resolve the apparent motion of the brightest emission knots after an elapsing time
of less than 10 years.

\section*{Acknowledgments}

We acknowledge the support from the LBT-Italian Coordination Facility for the 
execution of observations, data distribution and reduction.

%\clearpage

\clearpage

\clearpage

\clearpage

\clearpage

\clearpage

\clearpage

\bsp

\label{lastpage}


\clearpage



\begin{thebibliography}{99}
\bibitem[\protect\citeauthoryear{Barlow et al.}{1994}]{BA94} Barlow, M.J., Drew,J. E. Meaburn, J., Massey,R.M., 1994, MNRAS, 268, L29
\bibitem[\protect\citeauthoryear{Esposito et al.}{2012}]{ES12} Esposito, S. et al., 2012, in Adaptive Optics Systems III Proceedings of the SPIE, 8447, 84476B
\bibitem[\protect\citeauthoryear{Esposito et al.}{2013}]{ES13} Esposito, S. et al., 2013, A\&A, 549, A52
\bibitem[\protect\citeauthoryear{Guerra et al.}{2012}]{GU12} Guerra, J., Boutsia, K., Rakich, A., et al. 2012, PISCES Technical Report
\bibitem[\protect\citeauthoryear{Hill}{2010}]{HI10} Hill, J. M., 2010, Appl. Opt., 49, D115
\bibitem[\protect\citeauthoryear{Latherer \& Zickgraf}{1987}]{LA87} Latherer, C., Zickgraf, F.J., 1987, A\&A, 174, 103
\bibitem[\protect\citeauthoryear{McCarthy et al.}{2001}]{MC01} McCarthy et al., 2001, PASP, 113, 353
\bibitem[\protect\citeauthoryear{Meaburn et al.}{1996}]{ME96} Meaburn, J., Lopez, J. A., Barlow, M. J.,  Drew, J. E., 1996, MNRAS, 283, L69
\bibitem[\protect\citeauthoryear{Meaburn, Lopez \& O'Connor,}{1999}]{ME99} Meaburn, J., Lopez, J. A., O'Connor, J. A., 1999, ApJ, 516, L29
\bibitem[\protect\citeauthoryear{Meaburn et al.}{2004}]{ME04} Meaburn, J., Boumis, P., Redman, P., Lopez, J. A.,  Mavromatakis, F., 2004, A\&A, 422, 603
\bibitem[\protect\citeauthoryear{Najarro}{2001}]{NA01} Najarro, F., 2001, in ASP Conf. Ser. 233, P Cygni 2000: 400 Years of Progress, ed. M. de Groot \& C. Sterken (San Francisco, CA: ASP), 131
\bibitem[\protect\citeauthoryear{Najarro, Hillier \&  Stahl}{1997}]{NA97} Najarro, F., Hillier, D. J.,  Stahl, O., 1997, A\&A, 326, 1117
\bibitem[\protect\citeauthoryear{Nota, Livio, \& Clampin}{1995}]{NO95} Nota A., Livio M., Clampin M., 1995, ApJ, 448 788
\bibitem[\protect\citeauthoryear{Ragazzoni}{1996}]{RA96} Ragazzoni, R., 1996, Journal of Modern Optics, 43, 289
\bibitem[\protect\citeauthoryear{Riccardi et al.}{2010}]{RI10} Riccardi, A. et al., 2010, in Adaptive Optics Systems II Proceedings of the SPIE,  7736, 77362C
\bibitem[\protect\citeauthoryear{Salinari,  del Vecchio \& Biliotti}{1993}]{SA93} Salinari, P.,  del Vecchio, C., Biliotti, V., 1993, in Proceedings of the ICO-16 Conference on Active and Adaptive Optics, 247
\bibitem[\protect\citeauthoryear{Smith}{2001}]{SM01} Smith, N. 2001, in ASP Conf. Ser. 233, P Cygni 2000: 400 Years of Progress, ed. M. de Groot C. Sterken (San Francisco: ASP), 125
\bibitem[\protect\citeauthoryear{Smith \& Hartigan}{2006}]{SM06} Smith, N., Hartigan, P., 2006, ApJ, 638, 1045
\bibitem[\protect\citeauthoryear{White \& Becker}{1982}]{WB82} White, R. L., Becker, R.H., 1982, ApJ, 262, 657
\end{thebibliography}
\end{document}